\documentclass[twocolumn,nofootinbib,superscriptaddress]{revtex4-1}
\usepackage{latexsym,epsfig,amssymb, amsmath, nicefrac}
\usepackage{dsfont}
\usepackage[utf8]{inputenc}
\usepackage{amsmath}
\usepackage[top=1in, bottom=1in, left=1in, right=1in]{geometry}
\usepackage{color}

\newcommand{\bea}{\setlength\arraycolsep{2pt} \begin{eqnarray}}
\newcommand{\eea}{\end{eqnarray}}

\usepackage{hyperref}

\newsavebox{\uuunit}
\sbox{\uuunit}
{\setlength{\unitlength}{0.825em}
	\begin{picture}(0.6,0.7)
	\thinlines
	\put(0,0){\line(1,0){0.5}}
	\put(0.15,0){\line(0,1){0.7}}
	\put(0.35,0){\line(0,1){0.8}}
	\multiput(0.3,0.8)(-0.04,-0.02){12}{\rule{0.5pt}{0.5pt}}
	\end {picture}}

\def\be{\begin{equation}}
\def\ee{\end{equation}}
\def\ba{\begin{array}}
\def\ea{\end{array}}
\def\bea{\begin{eqnarray}}
\def\eea{\end{eqnarray}}
\def\bd{\begin{displaymath}}
\def\ed{\end{displaymath}}

\def\a{\alpha}
\def\b{\beta}

\def\f{\phi}

\def\l{\lambda}

\begin{document}

\vspace{1cm}

\title{Exorcising the Ostrogradsky ghost in coupled systems}

\author{Remko Klein}
\email{remko.klein@rug.nl}
\author{Diederik Roest}
\email{d.roest@rug.nl}
\affiliation{Van Swinderen Institute for Particle Physics and Gravity, University of Groningen, \\ Nijenborgh 4, 9747 AG Groningen, The Netherlands}

\begin{abstract}
The Ostrogradsky theorem implies that higher-derivative terms of a single mechanical variable are either trivial or lead to additional, ghost-like degrees of freedom. In this letter we systematically investigate how the introduction of additional variables can remedy this situation. Employing a Lagrangian analysis, we identify conditions on the Lagrangian to ensure the existence of primary and secondary constraints that together imply the absence of Ostrogradsky ghosts. We also show the implications of these conditions for the structure of the equations of motion as well as possible redefinitions of the variables.  We discuss applications to analogous higher-derivative field theories such as multi-Galileons and beyond Horndeski.

\end{abstract}	

\maketitle

\section{Introduction}

There has been a lot of recent interest in higher-derivative theories \cite{GLPV:beyond1,GLPV:beyond2,Zuma:transforming,Deffayet:counting,Langlois1,Langlois2,Langlois3,Langlois4,Gianmassimo1,Gianmassimo2}.  One of the reasons is their potential applicability to describe a wide range of physical phenomena. Moreover, the nature of these higher-derivative terms forces one to rethink the usual lore about higher-derivative Lagrangians and the occurance of Ostrogradsky ghosts. The Ostrogradsky theorem \cite{MO} tells us that a non-degenerate higher-derivative theory, is bound to contain ghost-like degrees of freedom. The only way to evade this is by considering degenerate theories \cite{Woodard,Chen}.

In field theories of such type, the appearance of the second derivative of a field does not necessarily imply its field equation to be fourth order in derivatives, and hence to have additional propagating degrees of freedom; instead, a very specific set of higher-derivative theories leads to second-order field equations. These are referred to as Lovelock gravity  for the metric tensor \cite{Lovelock} and (generalized) Galileons for a scalar field \cite{Rattazzi,Deffayet:generalized,Deffayet:k-essence}.

Yet more interesting are field theories that describe the couplings between these fields. Potential applications include inflation during the primordial Universe and quintessence during the late Universe. All possible couplings between a spin-2 and a spin-0 field, that lead to second-order field equations, are described by Horndeski's theory \cite{Horndeski, Deffayet:k-essence}. The above theories all have the property that the second time derivatives can always be eliminated via a (non Lorentz covariant) total derivative and hence enter the theory linearly.

Further generalizations have been proposed that go beyond Horndeski \cite{GLPV:beyond1} in the sense that not all second-order time derivatives can be eliminated via a total derivative. This implies that their field equations are in fact higher-order. A common feature of these generalizations is that the more complicated higher-derivative nature is constrained to the scalar sector of the theory, whilst the gravitational interactions are of the healthy Lovelock form\footnote{For vector instead of scalar fields, Galileon-like theories as well as their couplings to gravity have also been constructed, including the possibility to go beyond second-order field equations \cite{TasinatoVector,Heisenberg1,Tasinato,Rodriguez,Heisenberg2}.}. In addition, those theories that have been thoroughly analysed \cite{Langlois1,Langlois2,Langlois3,Gianmassimo1,Gianmassimo2} are restricted to up to quadratic dependence on second-order time derivatives.

To clarify in a more systematic way the possible structures in this kind of coupled theories, we focus on the simpler example of mechanical systems, lacking field-theoretic notions as spin and Poincare symmetry and/or diffeomeorphism invariance. The best-known class of such Lagrangians for $M$ variables $q_\a$ is
 \begin{align}
 L = L( q_\a, \dot q_\a) \,. \label{constraint-system}
  \end{align}
This system describes $M$ degrees of freedom provided the Jacobian $\partial^2 L / \partial \dot q_\a \partial \dot q_\b$ is non-singular. Similarly, in the case of a single variable, the Ostrogradsky theorem implies that the higher-derivative terms in
 \begin{align}
  L = L(\f, \dot \f, \ddot \f) \,, \label{higher-one}
 \end{align}
are either 1) trivial: they are linear and hence can be eliminated by means of a total derivative, or 2) fatal: the system has an additional, ghost-like degree of freedom. The field-theoretic extension of this idea includes the non-trivial interactions of Galileons and Lovelock gravity.

The mechanical counterpart to Horndeski and beyond is the merger of these simple mechanical systems. We will thus focus on the dynamics, constraints and degrees of freedom of the coupled system
 \begin{align}
  L =L( \f, \dot \f, \ddot \f ; q_\a, \dot q_\b) \,,
 \label{coupled-one}
 \end{align}
as well as its generalization to $N$ higher-order variables $\f_i$. To this end we will perform a systematic analysis employing a Lagrangian algorithm to derive the possible constraint structures. This results in very general conditions for systems such as \eqref{coupled-one} to be healthy, and gives us a handle on how to approach their field-theoretic counterparts.

The organization of this letter is as follows. The prerequisite Lagrangian analysis in terms of constraints is reviewed in section II. Subsequently, we apply this to the multi-variable generalization of \eqref{higher-one}. In Section IV we analyse the multi-variable extension of \eqref{coupled-one} and derive sufficient conditions for such theories to be free of Ostrogradsky ghosts. We also discuss in detail what these conditions imply for the structure of the equations of motion.  Section VI contains our conclusions and discusses possible generalizations and applications.

\medskip

\noindent
{\bf NOTE ADDED:} Upon preparation of this manuscript, we received the preprint \cite{Langlois4} employing a different method to analyze similar systems.

\section{Constraint analysis}

We first review the systematic and constructive algorithm to analyze the constraint structure of a given Lagrangian theory; for further details, see e.g.~\cite{Algorithm1,Algorithm2}. Our starting point will be the general Lagrangian \eqref{constraint-system}. Any higher-order theory can be cast into such a first-order form by the introduction of auxiliary variables. The above framework is therefore both necessary and sufficient for an analysis of all Lagrangian mechanical systems. 

For a generic Lagrangian, such a system does not have constraints and therefore describes $M$ propagating degrees of freedom. However, many interesting theories actually are constrained, such as all theories with gauge symmetries. The same applies to the present topic of healthy higher-order theories: their required degeneracy stems from constraints. In these cases, the algorithm that enables one to extract all the constraints from a given Lagrangian system consists of a number of identical consecutive steps.

The starting point of the analysis performed in each step is formed by a set of 'equations of motion'. In the first step these are the actual equations of motion following from the Lagrangian \eqref{constraint-system}. Subsequently, by forming linear combinations, one separates this set into second-order dynamical equations and (up to) first-order constraint equations. The time derivatives of the constraint equations are then added to the previous set of 'equations of motion' to form the starting point of the next step. This iterative process terminates when no new constraints are obtained in a given step: one has uncovered all the constraints within the system, allowing for an analysis of the dynamics of the propagating degrees of freedom.

In more detail, for the system above, the equations of motion in the first step are given by: \footnote{We denote partial derivatives of $L$ via subscripts, e.g. $$\partial^2 L/ \partial \dot q_\a \partial \dot q_\b =  L_{\dot{q}_\a \dot{q}_\b}$$}
\begin{align}
E^{(1)}_{\a} &= L_{\dot{q}_\a \dot{q}_\b}\ddot{q}_\b + L_{\dot{q}_\a q_\b}\dot{q}_\b - L_{q_\a} \notag \,, \\
&  \equiv W^{(1)}_{\a\b}\ddot{q}_\b + K^{(1)}_\a \,,
\end{align}
where both $W^{(1)}_{\a\b}$ and $K^{(1)}_\b$ depend on at most first derivatives of the coordinates. This system is constrained if some linear combination of the equations are in fact not second-order differential equations. This is equivalent to the existence of left null vectors of the Jacobian $W^{(1)}_{\a\b}$. 

To see this, let us assume that $m_1$ such null vectors exist and call them $v_a$ (with $a = 1,...,m_1 \leq M $). Then, we can take particular combinations of the equations of motion that at most depend on first derivatives, namely:
\begin{align}
C^{(1)}_a &= (v_a)_\a E^{(1)}_\a = (v_a)_\a K^{(1)}_\a \,.
\end{align}
These are the constraints corresponding to the null vectors.The internal structure of these constraints is as follows: 
\begin{itemize}
\item $g_1$ linear combinations vanish identically. This means that these combinations of the equations of motion vanish identically and thus hold off-shell; these are called gauge identities, as their existence is related to the presence of gauge symmetries. These will not play a role in the present discussion.
\item $a_1$ independent combinations do not depend on first derivatives of the coordinates, yielding algebraic constraints.
\item $d_1$ independent combinations depend on the first derivatives of the coordinates, yielding differential constraints.
\end{itemize}
Together all the types add up to $g_1+a_1+ d_1 = m_1$. Note that one must keep in mind that the Jacobian should be evaluated on the surface defined by the constraints. Doing so might lower the rank of $W^{(1)}_{\a\b}$, thus resulting in additional constraints and/or gauge identities. One should then repeat the analysis above, until no further decrease in rank occurs.

Assuming the analysis of the first step has been carried out in total, leading to $g_1$ gauge identities and $l_1 = a_1 + d_1$ primary constraints, one turns to the next step of the algorithm. Here one augments the original equations of motion with the time derivatives of the $d_1$ differential constraints $C^{(1)}_d$ and second-order time derivatives of the $a_1$ algebraic constraints $C^{(1)}_a$:
\begin{align}
E^{(2)}_A =  \begin{pmatrix} E^{(1)}_\a \\ \frac{d}{dt}C^{(1)}_d \\ \frac{d^2}{dt^2}C^{(1)}_a \end{pmatrix} = W^{(2)}_{A\b}\ddot{q}_\b + K^{(2)}_A ,
\end{align}
with $A = 1,\ldots,M + l_1 $. These are analysed in precisely the same manner as in the previous step. One in general ends up with an additional $g_2$ gauge identities and $l_2 = a_2 + d_2$ secondary constraints. Time evolving these new constraints and adjoining them to the previous set of equations provides the starting point for the next step. This process continues until at some point no new constraints (but possibly new gauge identities) are generated; one thus eventually ends up with a set of $l = l_1 + l_2 + l_3 + \ldots$ constraints and $g = g_1 + g_2 + g_3 + \ldots $ gauge identities.

After the termination of this algorithm, the number of degrees of freedom present in the theory can be computed to be \cite{DOFcount1,DOFcount2,DOFcount3,DOFcount4}
\begin{align}
M - \frac{1}{2}(l + g + e) \,.
\end{align}
Here $e$ is the total number of effective parameters (meaning parameters as well as their derivatives) describing the gauge transformations of the variables. These gauge transformations can be obtained from the corresponding gauge identities by noting that particular combinations of the generated gauge identities lead to Noether identities. These are, by the converse of Noether's second theorem, in one-to-one correspondence with gauge symmetries, which one can then explicitly obtain \cite{DOFcount3}.

The relation with the degree of freedom counting in the Hamiltonian formalism can be seen by noting that the number of second class Hamiltonian constraints is given by $l + g - e$ and the number of first class Hamiltonian constraints equals $e$. Also, the total number of gauge identities $g$ equals the number of primary first class constraints in the Hamiltonian analysis; see again ~\cite{DOFcount1,DOFcount2,DOFcount3,DOFcount4}.

\section{Higher-derivative sector}
We will now perform a detailed analysis of a theory solely involving $N$ second-derivative variables:
\begin{align}
L = L(\f_i,\dot{\f}_i,\ddot{\f}_i) \,.
\end{align}
Throughout our analysis we make two assumptions. Firstly, we treat all the variables on the same footing, meaning that constraints at a given step should always come in packages of $N$. Secondly, we assume the theory has no gauge symmetries, so that no gauge identities are generated throughout the analysis.

\subsection{Lagrangian analysis}

First of all, this theory can be put into a first order form by introducing auxiliary fields: 
\begin{align}
L(\f_i,A_i,\dot{A}_i) + \l^i(\dot{\f}_i - A_i) \,.
\end{align}
In the first step of the analysis we obtain
\begin{align}
W^{(1)} = \begin{pmatrix} L_{\dot{A}_i\dot{A}_j} & O & O \\
O & O & O \\
O & O & O \end{pmatrix}, \quad K^{(1)} = \begin{pmatrix} K_{A_i} \\ K_{\f_i} \\ K_{\l^i} \end{pmatrix} \,.
\end{align} 
It is immediate that two sets of $N$ constraints are generated due to the last two rows and are given by
\begin{align}
C_{\f_i} &= K_{\f_i} = \dot{\l}^i - L_{\phi_i}  \,, \notag \\
C_{\l^i} &=  K_{\l^i} = -(\dot{\f}_i- A_i) \,.
\end{align} 
These $2N$ 'auxiliary constraints' are solely due to the redundancy we put in by hand by introducing the auxiliary variables $A_i$ and their corresponding multipliers $\l^i$. Now, if $L_{\dot{A}_i\dot{A}_j}$ is non-degenerate, no further constraints are present. The 'auxiliary constraints' do not generate secondary constraints in the next step, so the analysis stops there and one finds that the theory propagates $3N - \frac{1}{2}(2N) = 2N$ degrees of freedom, as expected from a non-degenerate theory of $N$ second-derivative variables.\\

Keeping in mind our 'equal footing' assumption, we now consider the case where $N$ additional primary constraints are present. This is the case if and only if the \textit{primary condition}
\begin{align}
L_{\dot{A}_i\dot{A}_j} = 0 \,,
\end{align}
holds, in which case the additional constraints are given by
\begin{align}
C_{i} &= K_{A_i} = L_{\dot{A}_i A_j}\dot{A}_j + L_{\dot{A}_i \f_j}\dot{\f}_j - L_{A_i} + \lambda^i \,.
\end{align}
The $3N$ primary constraints are not sufficient to reduce the number of degrees of freedom to $N$; rather we need an additional $N$ secondary constraints. Taking the time derivatives of our constraints, we proceed to step 2 and obtain
\begin{align}
W^{(2)} = \begin{pmatrix} O & O & O \\
O & O & O \\
O & O & O \\
Z_{ij} & \tilde{Z}_{ij} & O \\
-L_{\f_i\dot{A}_j} & O & \mathds{1} \\
O & -\mathds{1} & O
\end{pmatrix}, \quad K^{(2)} = \begin{pmatrix} K_{A_i} \\ K_{\f_i} \\ K_{\l^i} \\ K^{(2)}_{C_i} \\ K^{(2)}_{\f_i} \\ K^{(2)}_{\l^i} \end{pmatrix} \,,
\end{align}
using the notation
\begin{align}
\frac{d}{dt}C^{(1)}_i & \equiv Z_{ij}\ddot{A}_j +  \tilde{Z}_{ij}\ddot{\f}_j + K^{(2)}_{C_i} \,, \notag \\ Z_{ij} & = L_{\dot{A}_i A_j} - L_{A_i \dot{A}_j} \,.
\end{align}
Therefore there are $N$ secondary constraints if and only if the \textit{secondary condition} 
\begin{align}
L_{\dot{A}_i A_j} - L_{A_i \dot{A}_j} = 0 \,,
\end{align}
holds, in which case they are given by:
\begin{align}
D_i &= K^{(2)}_{C_i} + \tilde{Z}_{ij}K^{(2)}_{\l^j}  \,.
\end{align}
In the single-variable case the secondary condition is automatically satisfied: any theory satisfying the primary condition propagates $3 -\frac{1}{2}(3 + 1) = 1$ degree of freedom and is hence free of the Ostrogradsky ghost. In there is more than one variable ($N>1$) however, the primary condition does not imply the secondary condition, and is hence not sufficient to ensure the absence of Ostrogradsky ghosts \cite{Motohashi:2014opa}. If in addition the secondary condition holds, the theory propagates the healthy amount of $3N - \frac{1}{2}(3N + N) = N$ degrees of freedom.

If further constraints (tertiary, quaternary, etc.) are generated at subsequent steps, the number of degrees of freedom will be less than $N$. We will not consider this non-generic case.

\subsection{Equations of motion}
The primary and secondary conditions, written in terms of the original variables (by identifying $A_i$ with $\dot{\phi}_i$), are found to have a clear interpretation when looking at the equations of motion of the original higher-order formulation of the theory. These equations of motion are
\begin{align}
E_{\phi_i} = L_{\ddot{\f}_i\ddot{\f}_j}\phi_j^{(4)} + \textup{lower order terms} \,.
\end{align}
Therefore if $L_{\ddot{\f}_i\ddot{\f}_j}$ is invertible one can solve for all fourth-order derivatives and one obtains a set of $N$ genuinely fourth-order differential equations. Hence $4N$ initial conditions need to be specified and the theory describes $2N$ degrees of freedom, of which $N$ are Ostrogradsky ghosts.\\

One sees that if and only if the primary condition is satisfied, all the terms involving fourth-order derivatives vanish. Generically however, third-order derivatives are still present:
\begin{align}
E_{\phi_i} = (L_{\ddot{\f}_i \dot{\f}_j} - L_{\dot{\f}_i \ddot{\f}_j})\f_j^{(3)} + \textup{lower order terms} \,.
\end{align}
Then, if $(L_{\ddot{\f}_i \dot{\f}_j} - L_{\dot{\f}_i \ddot{\f}_j})$ is invertible one can solve for the third derivatives and one has $N$ third-order differential equations requiring $3N$ initial conditions, thus signalling Ostrogradsky ghosts still. Therefore, if and only if the primary \textit{and} secondary conditions are satisfied, both the fourth- and third-order terms vanish and one obtains $N$ second-order equations of motion describing $N$ degrees of freedom.

The above is also immediate by first observing that the primary condition implies that the second derivatives enter the Lagrangian linearly, i.e.
\begin{align}
L(\f_i,\dot{\f}_i, \ddot{\f}_i) = \ddot{\f}_i f^{i}(\f_i,\dot{\f}_i) + g(\f_i,\dot{\f}_i) \,.
\end{align}
In the single-variable case this is sufficient to conclude that the second derivative enters only through a total derivative, i.e.
\begin{align}
L(\f,\dot{\f}, \ddot{\f}) = \frac{d}{dt}F(\f,\dot{\f}) - F_{\f}(\f,\dot{\f})\dot{\f} + g(\f,\dot{\f}) \,,
\end{align}
with $F_{\dot{\f}} = f$.
In the multi-variable case one in addition needs the secondary condition, which implies that $f^i_{\dot{\f}_j} = f^j_{\dot{\f}_i}$. This is a necessary and sufficient condition for the existence of an $F(\f_i ,\dot{\f}_j)$ such that $F_{\dot{\f}_i} = f^i$, thus implying the following form for the Lagrangian 
\begin{align}
L(\f_i,\dot{\f}_i, \ddot{\f}_i) = \frac{d}{dt}F(\f_i,\dot{\f}_i) - F_{\f_j}(\f_i,\dot{\f}_i)\dot{\f}_j  + g(\f_i,\dot{\f}_i)\,.
\end{align}
The extension of the higher-derivative sector therefore does not change the original conclusion: the higher-derivative terms are either trivial or fatal.

\section{Adding a healthy sector}

\noindent
We will now consider the general case of $N$ problematic and $M$ healthy, first-order variables:
$$ L(\phi_i,\dot{\phi}_i,\ddot{\phi}_i,q_\alpha,\dot{q}_\alpha) \,,$$
with $i = 1,...,N$ and $\a = 1,...,M $. We will again work under the same assumptions as in the previous section. In addition we also exclude possible degeneracies in the healthy sector, which amounts to demanding that the Jacobian $L_{\dot{q}_\a \dot{q}_\b}$ is non-degenerate (i.e. invertible).

\subsection{Lagrangian analysis}

Again, this can be brought to the first-order form \eqref{constraint-system} by the introduction of Lagrange multipliers:
\begin{align}
L(\phi_i,A_i,\dot{A}_i,q_\alpha,\dot{q}_\alpha) + \lambda^i (\dot{\phi}_i -A_i) \,.
\end{align}
In the first step, this theory has the Jacobian:
\begin{align}
W^{(1)} = \begin{pmatrix}
L_{\dot{A}_i\dot{A}_j} & L_{\dot{A}_i \dot{q}_\b}  & O  & O \\ 
L_{\dot{q}_\alpha \dot{A}_j }&L_{\dot{q}_\alpha \dot{q}_\beta}  & O &O \\ 
O& O &O  &O \\ 
O&  O&  O& O
\end{pmatrix}\,.
\end{align}
Again, we find the 'auxiliary constraints' resulting from the latter two rows; these will not generate secondary constraints. Thus if there are no additional constraints, i.e. if rank$(W^{(1)}) = N+M$, then the theory describes $2N +M$ degrees of freedom of which $N$ are Ostrogradsky ghosts.

For this reason, we will assume an additional $N$ primary constraints are present implying the existence of $N$ additional null vectors, which can be chosen as
\begin{align}
V_i = (\mathds{1}_{ij},V_{i\alpha},O,O) \,.
\end{align}
This follows straightforwardly from the invertibility of $L_{\dot{q}_\alpha \dot{q}_\beta}$. Writing out the defining properties of these eigenvectors, i.e.
\begin{align}
L_{\dot{A}_i\dot{A}_j} + V_{i\a}L_{\dot{q}_\alpha \dot{A}_j } = L_{\dot{A}_i \dot{q}_\b} + V_{i\a}L_{\dot{q}_\alpha \dot{q}_\beta} = 0 \,,
\end{align}
and again using the invertibility of $L_{\dot{q}_\alpha \dot{q}_\beta}$, we obtain:
\begin{align}
V_{i\a} = - L_{\dot{A}_i\dot{q}_\b}(L_{\dot{q}_\b \dot{q}_\a})^{-1} \,.
\end{align}
Consistency of these equations requires
\begin{align}
L_{\dot{A}_i\dot{A}_j} = L_{\dot{A}_i \dot{q}_\a}(L_{\dot{q}_\a \dot{q}_\b})^{-1}L_{\dot{q}_\b \dot{A}_j} \,, \label{primary-conditions}
\end{align}
which is our \textit{primary condition} equivalent to the existence of the $N$ null vectors and hence $N$ primary constraints. We thus end up with the following $3N$-constraints:
\begin{align}
C^{(1)}_{\phi_i} &= K_{\phi_i} = \dot{\l}^i - L_{\phi_i} \,, \notag \\
C^{(1)}_{\lambda^i} &= K_{\lambda^i} = -(\dot{\phi}_i - A_i) \,, \notag \\
C^{(1)}_{i} &= K_{A_i} + V_{i\a}K_{q_\a} \,.
\end{align}
Time-evolving these yields the augmented Jacobian
$$ W^{(2)} = \begin{pmatrix}
L_{\dot{A}_i\dot{A}_j} & L_{\dot{A}_i \dot{q}_\alpha}  & O  & O \\ 
L_{ \dot{q}_\alpha\dot{A}_i}&L_{\dot{q}_\alpha \dot{q}_\beta}  & O &O \\ 
O& O &O  &O \\ 
O&  O&  O& O \\
Z_{ij} & Z_{i\alpha}  & \tilde{Z}_{ij} & O \\ 
-L_{\phi_i\dot{A}_j} & -L_{\phi_i \dot{q}_\alpha}  &O  &\mathds{1} \\ 
O& O &-\mathds{1}   & O
\end{pmatrix} \,,$$
where now 
\begin{align}
\frac{dC_i}{dt} \equiv Z_{ij}\ddot{A}_j + Z_{i\alpha}\ddot{q}_\alpha +\tilde{Z}_{ij}\ddot{\f}_j +  K^{(2)}_{C_i}  \,.
\end{align}

We will assume there are $N$ additional secondary constraints. This implies the existence of $N$ eigenvectors $X_{i}$ that are linearly independent from each other and the first generation eigenvectors.  We can take them to be of the form:
\begin{align}
X_i = (O,X_{i\a},O,O,\mathds{1}_{ij},O,\tilde{Z}_{ij}) \,.
\end{align}
This then implies the relations
\begin{align}
Z_{ij} + X_{i\alpha}L_{\dot{q}_\alpha \dot{A}_j} = Z_{i\alpha} + X_{i\beta}L_{\dot{q}_\beta\dot{q}_\alpha} = 0 \,.
\end{align}
Using invertibility, one can solve for the form of the null eigenvectors:
\begin{align}
X_{i\a} = - Z_{i\b}(L_{\dot{q}_\b \dot{q}_\a})^{-1} \,,
\end{align}
and one obtains a necessary and sufficient condition for the existence of these null vectors, namely
\begin{align}
Z_{ij} &= Z_{i\a}(L_{\dot{q}_\a \dot{q}_\b})^{-1}L_{\dot{q}_\b \dot{A}_j} \,,
\end{align}
which is therefore equivalent with the existence of $N$ secondary constraints. In terms of the Lagrangian, this \textit{secondary condition} reads\footnote{In principle since $Z_{ij}$ and $Z_{i\a}$ contain third-order partial derivatives of the Lagrangian, one would expect that the secondary condition should too. However, these terms enter only as partial derivatives of the primary condition, and hence vanish upon imposing it. Explicitly these particular combinations are of the form
	\begin{align}
	L_{\dot{A}_i\dot{A}_j \psi} + V_{i\alpha}L_{\dot{q}_\a\dot{A}_j \psi} + L_{\dot{A}_i\dot{q}_\a \psi}V_{\a j} + V_{i\a}L_{\dot{q}_\a\dot{q}_\b \psi} V_{\b j}\,,  \notag 
	\end{align}
	where $\psi$ is an arbitrary variable.}
\begin{align}
L_{A_i \dot{A}_j}   - L_{\dot{A}_i A_j} = & V_{i\a} (L_{\dot{q}_\a A_j} - L_{q_\a\dot{A}_j})  \notag \\
& + (L_{\dot{A}_i q_\a} - L_{A_i \dot{q}_\a})V_{\a j}  \notag \\
&  + V_{i\a}(L_{\dot{q}_\a q_\b} - L_{q_\a \dot{q}_\b})V_{\b j} \,. \label{secondary-conditions}
\end{align}
The generated secondary constraints are given by:
\begin{align}
D^{(2)}_i &= X_{i\alpha}K_{q_\alpha} + K^{(2)}_{C_i} + \tilde{Z}_{ij}\dot{A}_j \,.
\end{align}
Now we find that the time derivatives of our secondary constraints yield the further augmented Jacobian:
\begin{align}
W^{(3)} =  \begin{pmatrix}
L_{\dot{A}_i\dot{A}_j} & L_{\dot{A}_i \dot{q}_\alpha}  & O  & O \\ 
L_{ \dot{q}_\alpha\dot{A}_i}&L_{\dot{q}_\alpha \dot{q}_\beta}  & O &O \\ 
O& O &O  &O \\ 
O&  O&  O& O \\
Z_{ij} & Z_{i\alpha}  & \tilde{Z}_{ij} & O \\ 
-L_{\phi_i\dot{A}_j} & -L_{\phi_i \dot{q}_\alpha}  &O  &\mathds{1} \\ 
O& O &-\mathds{1}   & O \\
Y_{ij} & Y_{i\alpha} & \tilde{Y}_{ij} & \mathds{1}
\end{pmatrix} \,.
\end{align}
Since we have $N$ primary and $N$ secondary constraints (in addition to the trivial $2N$ auxiliary ones) we will assume that there are no further null eigenvectors. In that case the iterative process stops here. 

We therefore conclude that under the conditions \eqref{primary-conditions} and \eqref{secondary-conditions}, corresponding to the existence of $N$ primary and secondary constraints, we have $3N + M - \frac{1}{2}(3N + N) = N+ M$ propagating degrees of freedom.

\subsection{Equations of motion}

The equations of motion of a theory satisfying the constraints of the analysis above, are of a very special kind: they are higher order but can be rewritten into second-order equations. To see this explicitly for the higher-derivative sector, let us consider the original equations of motion. These can be written as 
\begin{align}
E_{\phi_i} & = \frac{d}{dt}(L_{\ddot{\phi}_i\ddot{\phi}_j} \dddot{\phi}_j + L_{\ddot{\phi}_i \dot{q}_\alpha}\ddot{q}_\alpha) + \frac{d}{dt}K_{\dot{\phi}_i} - K_{\phi_i} \,, \notag \\
E_{q_\alpha} & = L_{\dot{q}_\alpha\ddot{\phi}_j} \dddot{\phi}_j + L_{\dot{q}_\alpha\dot{q}_\beta}\ddot{q}_\beta + K_{q_\alpha} \,.
\end{align}
They are fourth/third and third/second order in derivatives of $\phi_i$/$q_\a$, respectively. Based on the constraint analysis of the previous section, we consider the following combinations
\begin{align}
\tilde{E}_{\phi_i} = E_{\phi_i} + \frac{d}{dt}(V_{i\alpha}E_{q_\alpha}) \,.
\end{align}
By virtue of the primary condition \eqref{primary-conditions}, the highest derivative terms drop out, resulting in
\begin{align}
\tilde{E}_{\phi_i} =& Z_{ij}\dddot{\phi}_j + Z_{i\a}\ddot{q}_\alpha + \tilde{Z}_{ij}\ddot{\f}_j +  K_{C_i}^{(2)}- K_{\phi_i} \,,
\end{align}
which are third/second order. Next one takes the further combinations,
\begin{align}
\bar{E}_{\f_i} = \tilde{E}_{\f_i} + X_{i\a}E_{q_\a} \,,
\end{align}
which are seen to be of second/first order if the secondary condition holds. The new combinations are then given by
\begin{align}
\bar{E}_{\f_i} = D^{(2)}_i - K_{\f_i} \,.
\end{align}
As a last step we would like to eliminate the third-order derivatives in the $M$ equations $E_{q_\a}$ via the time derivatives of $\bar{E}_{\f_i}$. {To this end, one can consider the following combinations}
\begin{align}
\bar{E}_{q_\a} &=  E_{q_\a} + B_{\a i}\frac{d}{dt}\bar{E}_{\phi_i} \notag \\
& = L_{ \dot{q}_\alpha\dot{A}_j}\dddot{\f}_j + B_{\a i}(Y_{ij} + L_{\f_i \ddot{\f}_j})\dddot{\f}_j + \ldots \,.
\end{align}
The third-order derivatives drop out precisely if one can find $M$ vectors $B_\a$ (not necessarily distinct or nonzero) such that, 
\begin{align}
L_{ \dot{q}_\alpha\dot{A}_j} + B_{\a i}(Y_{ij} + L_{\f_i \ddot{\f}_j}) = 0 \,.
\end{align} 
The end result will then be a set of $N + M$ second-order differential equations, $\bar{E}_{\f_i}$ and $\bar{E}_{q_\a}$, equivalent to the original equations of motion. Therefore one needs to specify $2N + 2M$ initial conditions and the theory describes $N+M$ degrees of freedom (not considering non-generic cases with additional constraints).

\section{Variable redefinitions}

Given a healthy higher-derivative theory as presented in the previous section, one should wonder whether it is not simply a manifestly healthy theory written in terms of poorly chosen variables. In this section we will derive necessary conditions for this to be possible.

We will first consider the case of a single higher-derivative variable and a single healthy variable, i.e. $L(\f,\dot{\f},\ddot{\f},q,\dot{q})$. Assume that, up to a total derivative, the theory can be recast into a manifestly healthy form by means of an invertible redefinition of the variables. The theory in terms of the redefined variables should depend on at most first derivatives and should be non-degenerate. The most general relation is then\footnote{Dependence on higher than first-order derivatives of $\f$ in both the function $f$ as well as $\bar{q}$ are inconsistent with the general dependence of $L$ and $\bar{L}$ combined with $L_{\dot{q}\dot{q}} \neq 0$. }
\begin{align}
&L(\f,\dot{\f},\ddot{\f},q,\dot{q}) + \frac{d}{dt}f(\f,\dot{\f},q) = \bar{L}(\phi,\dot{\phi},\bar{q},\dot{\bar{q}}) \,, \notag \\
&\bar{q} = \bar{q}(\f,\dot{\f},q),\quad \bar{q}_q \neq 0 \,.
\end{align}
One can easily verify that the theory is degenerate  by observing that
\begin{align}
L_{\ddot{\f}\dot{q}} &= \bar{L}_{\dot{\bar{q}}\dot{\bar{q}}} \bar{q}_{\dot{\f}}\bar{q}_{q} \,, \notag \\
L_{\dot{q}\dot{q}} &= \bar{L}_{\dot{\bar{q}}\dot{\bar{q}}} \bar{q}_{q} \bar{q}_{q} \,, \notag \\
L_{\ddot{\f}\ddot{\f}} &= \bar{L}_{\dot{\bar{q}}\dot{\bar{q}}}\bar{q}_{\dot{\f}}\bar{q}_{\dot{\f}} \,,
\end{align}
which  implies that the primary and secondary conditions are satisfied. Here we have used that the derivative of the new variable is given by
\begin{align}
  \dot{\bar{q}} = \bar{q}_{\dot{\f}}\ddot{\f} + \bar{q}_{\f}\dot{\f} + \bar{q}_{q}\dot{q} \,.
\end{align} 
Now the null vector characterizing the primary constraint, $V = (1,v)$, is such that
\begin{align}
v = - \frac{L_{\ddot{\f}\dot{q}}}{L_{\dot{q}\dot{q}}} = - \frac{\bar{q}_{\dot{\f}}}{\bar{q}_q} = v(\f,\dot{\f},q) \,.
\end{align}
Thus we find that the existence of a redefinition implies that $v$ depends only on $\dot{\f}$, $\f$ and $q$. In contrast, theories for which $v$ depends on $\ddot{\f}$ or $\dot{q}$ do not admit such a redefinition. An example is provided by 
\begin{align}
L(\f,\dot{\f},\ddot{\f},q,\dot{q}) = \frac{1}{2}\dot{\f}^2 + \frac{1}{2}\frac{\dot{q}^2}{1+ \ddot{\f}}  \,,
\end{align}
which was analysed in detail in \cite{Example}.

We can do a similar analysis for the general case involving several variables. To this end we consider the following relation
\begin{align}
&L(\f_i,\dot{\f}_i,\ddot{\f}_i,q_
\a,\dot{q}_\a) + \frac{d}{dt}f(\f_i,\dot{\f}_i,q_\a) = \bar{L}(\f_i,\dot{\f}_i,\bar{q}_\a,\dot{\bar{q}}_\a) \,, \notag\\
&\bar{q}_\a = \bar{q}_\a(\f_i,\dot{\f}_i,q_\b),\quad \textup{det}\frac{\partial \bar{q}_\a}{\partial q_\b} \neq 0 \,.
\end{align}
Again, one can easily verify that the theory is degenerate and has $N$ null vectors, $V_i = (\mathds{1}_{ij},V_{i\a})$, with 
\begin{align}
V_{i\a} &= - L_{\ddot{\f}_i \dot{q}_\b}(L_{\dot{q}_\b\dot{q}_\a})^{-1} \,, \notag \\
& = \frac{\partial \bar{q}_\b}{\partial \dot{\f}_i}(\frac{\partial \bar{q}_\b}{\partial q_\a})^{-1} = V_{i\a}(\f_j,\dot{\f}_j,q_\b) \,. \label{dependence}
\end{align}
Therefore $V_{i\a} = V_{i\a}(\f_j,\dot{\f}_j,q_\b)$ is a necessary condition for the existence of a healthy redefinition.

\section{Conclusions}

In this letter we have presented a systematic analysis of extensions of single-variable systems with higher derivatives. 

In the case of a higher-derivative sector with $N$ variables, the existence of $N$ primary constraints is equivalent to the vanishing of the Jacobian $L_{\ddot \phi_{i} \ddot \phi_j}$. In the single variable case this is equivalent to the requirement of linearity in the second derivative of the variable, which suffices to eliminate both the fourth- as well as third-order terms in the equation of motion. In contrast, in the case of multiple variables, one has to impose a second condition:  the anti-symmetric part of $L_{ \ddot \phi_{i}  \dot \phi_j}$ must be vanishing. This leads to the required $N$ secondary constraints to have the correct number of degrees of freedom. In this case all the higher-derivative terms combine into a total derivative.

This conclusion changes when augmenting this system with a healthy sector of $M$ variables $q_\alpha$. The conditions for the constraints' existence are in this case \eqref{primary-conditions} and \eqref{secondary-conditions} and allow the Jacobian and anti-symmetric part of the above matrix to be non-vanishing and to have a specific relation to the healthy sector. In such coupled systems, it is therefore indeed possible to {be free of} the Ostrogradsky ghosts while retaining non-trivial higher-derivative interactions. 

Interestingly, these  systems cannot always be brought to a manifestly first-order Lagrangian by means of a variable redefinition. A necessary condition for this to be possible imposes restrictions on the null vectors that generate the primary constraints: these are only allowed to have the specific dependence \eqref{dependence}. More general systems do not allow for such redefinitions and can nevertheless have healthy dynamics.

As a follow-up investigation, one could perform a similar analysis as presented in this paper for yet higher-derivative terms, e.g.~starting from terms involving cubic or quartic derivatives. While in the single-variable case  this again effectively amounts to a total derivative, it might be possible to remedy the Ostrogradsky instability in more general ways in the kind of coupled systems investigated here. In addition one could do a systematic analysis also allowing for the possibility of gauge symmetries.

Finally, it would be very interesting and phenomenologically relevant to extend this classical mechanics analysis to field theories. A systematic analysis along the lines of section II of the possible theories beyond Horndeski might lead to new possibilities involving second derivatives of the metric and a scalar. Also, adding additional scalars hence entering the domain of multi-Galileons \cite{Multi1,Multi2} and their covariantizations \cite{Covariantmulti1,Covariantmulti2} would be very interesting. In addition, one could try to construct field theories based on third and/or higher derivatives. This would possibly lead to higher-order variants of Lovelock gravity and Galileons. We leave these possibilities for future study.

\section*{Acknowledgments}
RK acknowledges the Dutch funding agency `Foundation for Fundamental Research on Matter' (FOM) for financial support.

\providecommand{\href}[2]{#2}\begingroup\raggedright\endgroup

\end{document}